\documentclass[%
 reprint,
superscriptaddress,
showpacs,
 amsmath,amssymb,
 aps,
]{revtex4-1}

\usepackage{graphicx}
\usepackage{dcolumn}
\usepackage{bm}
\usepackage{subfig}
\usepackage{float}
\usepackage{todonotes}
\usepackage{enumerate}

\def\ii{{\rm i}}
\begin{document}

\preprint{APS/123-QED}

\title{Large density stratification stabilizes Rayleigh--Taylor instability in presence of shear}
\author{Raunak Raj}


\author{Anirban Guha}
 \email{anirbanguha.ubc@gmail.com}
\affiliation{
Environmental and Geophysical Fluids Group, Department of Mechanical Engineering, Indian Institute of Technology Kanpur, U.P. 208016, India.\\}%


\date{\today}

\begin{abstract}
This letter investigates the effect of shear on Rayleigh--Taylor instability (RTI). Even simple uniform shear strongly influences the instability; longer waves are completely stabilized when  density stratification is large (higher Atwood numbers). This apparently counter-intuitive result is due to the presence of Atwood number in the shear term. When the unstable density interface is embedded in a  shear layer,  shear is again found to stabilize the RTI. However, this configuration introduces additional unstable shear instability modes in the lower wavenumber regime. A new type of shear instability, whose growth rate increases with Atwood number, plays a dominant role, while Kelvin--Helmholtz instability (KHI), which was previously understood to be the only possible shear instability in this context, has little significance.  Hence the billows observed in the nonlinear stages of RTI, which are usually attributed to KHI, may actually be the nonlinear manifestation of this new instability.

\end{abstract}


\maketitle



 Rayleigh--Taylor instability (RTI) is a familiar gravity driven flow phenomenon observed when a fluid of density $\rho_1$  rests on the top of a lighter one of density $\rho_2$  \citep{taylor1950instability}. RTI has varied applications in  industrial (e.g.\ aerosol transport, thin film flows, pool boiling, inertial confinement fusion), oceanic, and even in astrophysical flows.  In RTI, with gravity $g>0$, the system is unstable for Atwood number, $A_t \equiv (\rho_1-\rho_2)/(\rho_1+\rho_2)>0$ and for all wavenumber, $k$. Furthermore, the growth rate, $\gamma$, varies as $\sqrt{A_tgk}$, implying that shorter wavelengths are more unstable. Finding ways to stabilize the  explosive growth of RTI is of key interest in many sub-fields of fluids and plasmas. In presence of viscosity and/or surface tension, the short waves are known to be stabilized  \citep{chandrasekhar2013hydrodynamic,charru2011hydrodynamic}. Stabilization of RTI by Coriolis force has also been reported  \cite{tao2013nonlinear,baldwin2015inhibition}. For flows of geophysical and industrial relevance, the effect of shear on RTI is important. For example, in displacement flows  \citep{taghavi2010influence}, knowing the stability effect of shear on the unstable interface would be important for the oil industry.  Previous studies on the effect of shear on RTI reveal contradictory results. In one such study by  ref.\  \cite{guzdar1982influence}, who used continuous velocity profile for shear, it was concluded that the RTI growth rate is inhibited by the application of shear. However, treating shear as a discrete velocity jump across the density interface, it was shown in ref.\  \cite{zhang2005effect} that presence of shear destabilizes the instability even further. It was also argued that for high values of shear, the instability is mainly driven by shear and moreover, increasing the Atwood number stabilizes the flow, as is the case in a pure Kelvin--Helmholtz instability. 

The objective of this letter is the resolution of this apparent paradox by theoretically investigating the effect of background shear on RTI using  continuous, piecewise linear velocity profiles. Initially we assume nothing but a uniform shear across the density interface and then we make the setting more realistic by assuming a finite shear layer. Use of a broken line velocity profile helps in underpinning the cause of instabilities in terms of localized wave interactions. Additionally, we do this in a fully non-Boussinesq setting so as to incorporate the variation of density in the inertial terms as well. The Boussinesq approximation is severely limited by the fact that any variation of density in the terms other than the gravitational force term is ignored. This effectively means that the terms `density variation' and `buoyancy variation' are qualitatively treated as one and the same. While this approximation works reasonably well for lower values of $A_t$ in the absence of shear, for higher $A_t$ or even for high values of shear, there is a substantial qualitative as well as quantitative variation between the Boussinesq and the non-Boussinesq settings.



\begin{figure}
\centering\includegraphics[width=80mm]{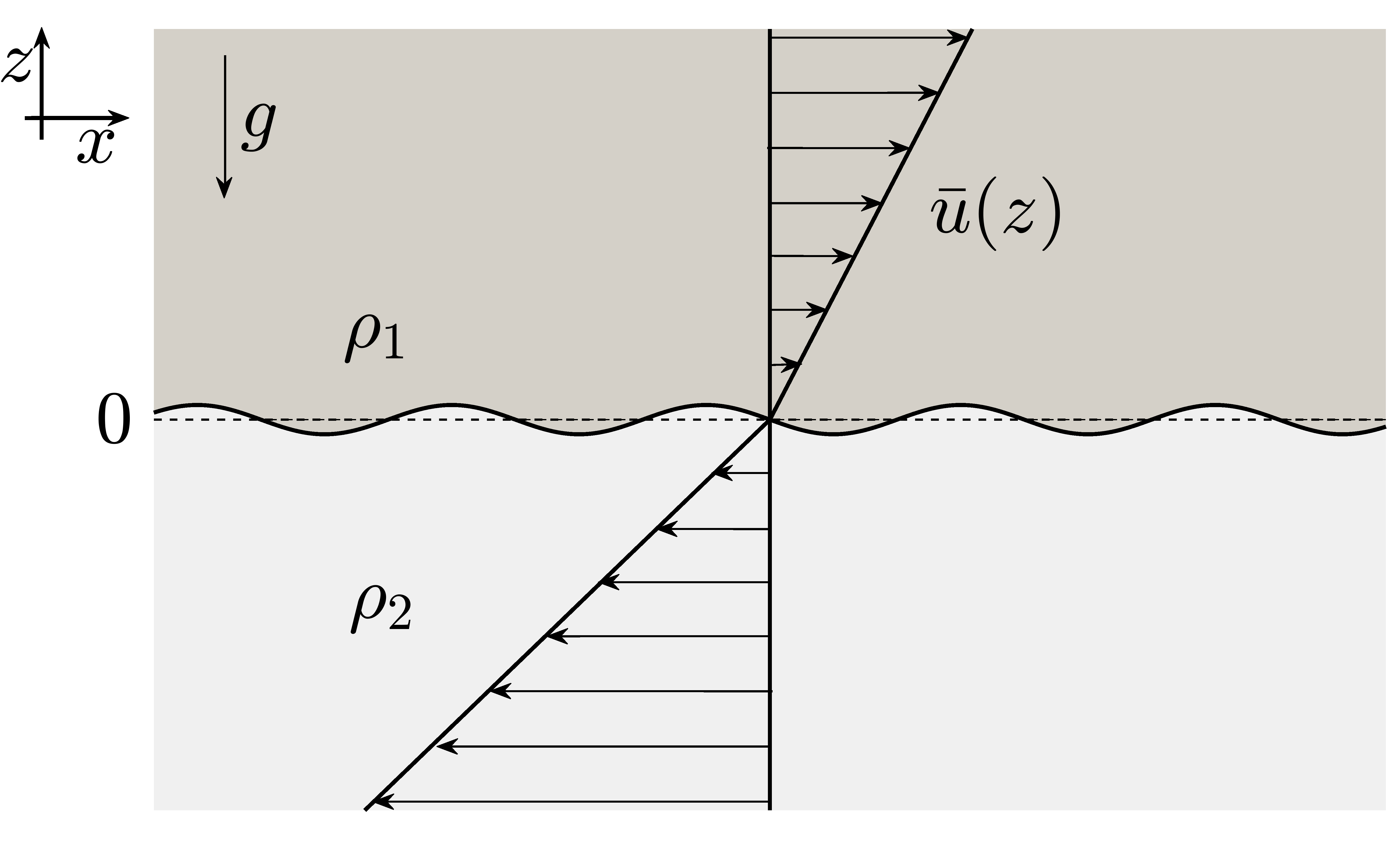}
  \caption{ Schematic of adversely stratified density interface in presence of piecewise uniform shear.}
  \label{fig:1}
\end{figure}

We consider a 2D flow in the $x-z$ plane having a piecewise uniform shear $\Omega(z)\equiv d\bar{u}/dz$, where $\bar{u}$ is the base velocity, imposed over an unstable density interface (see fig.\ \ref{fig:1}):
\begin{equation}
\bar{\rho}(z) = \left\{
        \begin{array}{cc}
        \rho_1 & \quad 0 < z \\
        \rho_2 & \quad  z < 0
        \end{array}
    \right.
   \qquad \Omega(z) = \left\{
        \begin{array}{cc}
        \Omega_1& \quad 0 < z \\
        \Omega_2 & \quad  z<0.
        \end{array}
    \right. 
\end{equation}
When $\Omega_1=\Omega_2$, the situation will be that of a uniform shear, but to keep our setting general, we don't invoke this assumption. Besides, it might be noted that even though we have vorticity present in the base flow, the perturbed flow is still irrotational everywhere except at the interface because both shear and  density are constant in each layer. For understanding this, we appeal to the linearized vorticity transport equation according to which, the total vorticity $\zeta$ is conserved in the fluid bulk for inviscid 2D flows. 
Baroclinic generation of vorticity is  restricted to the density interface(s) only. Thus, everywhere except at the interface, we have $D\zeta/Dt=0$.
Since our focus is on linear instabilities, the operator  $\bar{D}/\bar{D}t\equiv \partial/\partial t + \bar{u}\, \partial/\partial x$ denotes the \emph{linearized} material derivative. Writing $\zeta(x,z,t)=\Omega(z)+\zeta'(x,z,t)$ (decomposing into background and perturbation), we  obtain

\begin{align}
\frac{\bar{D}\zeta'}{\bar{D}t}\quad &=-w'\frac{d\Omega}{d z}.
\end{align}
In presence of a piecewise uniform shear, the R.H.S. is identically zero everywhere except at the interfaces. Therefore, the perturbation vorticity is zero within the bulk of the fluid, i.e. except at the interfaces and hence, we can still use the potential formulation for the perturbed flow. We introduce potentials $\phi_1$ and $\phi_2$ in the region above and below the interface (having equation $z=\eta(x,t)$) respectively, satisfying the incompressible continuity equation
\begin{align}
\nabla^2\phi_1=0 \qquad & z>0\\ 
\nabla^2\phi_2=0 \qquad & z<0
\label{eq:Lap1}.
\end{align}
We write the usual linearized kinematic boundary conditions just above and below the interface at $z=0$ as
\begin{equation}\label{eq:KBC}
\frac{\partial \eta}{\partial t}=\left.\frac{\partial \phi_1}{\partial z} \right\rvert_{z=0}\qquad ;\qquad\frac{\partial \eta}{\partial t}=\left.\frac{\partial \phi_2}{\partial z} \right|_{z=0}.
\end{equation}
The linearized dynamic boundary condition, which is obtained by integrating the momentum equation along the interface and matching the pressure across it, is
\begin{equation}\label{eq:DBC}
\rho_1 \left[ \frac{\partial \phi_1}{\partial t}-\Omega_1\psi+g\eta\right]_{z=0}=\rho_2 \left[ \frac{\partial \phi_2}{\partial t}-\Omega_2\psi+g\eta\right]_{z=0}.
\end{equation}
Here, $\psi$ is the perturbation streamfunction, and apart from the presence of `$\Omega\psi$' term, this equation is similar to the linearized Bernoulli's equation which we would have obtained in the absence of base shear. It might be noted that the convective term of the momentum equation, which gives rise to this term, cannot be integrated had the base shear not been piecewise constant. 

\begin{figure}
\centering
\includegraphics[width=80mm]{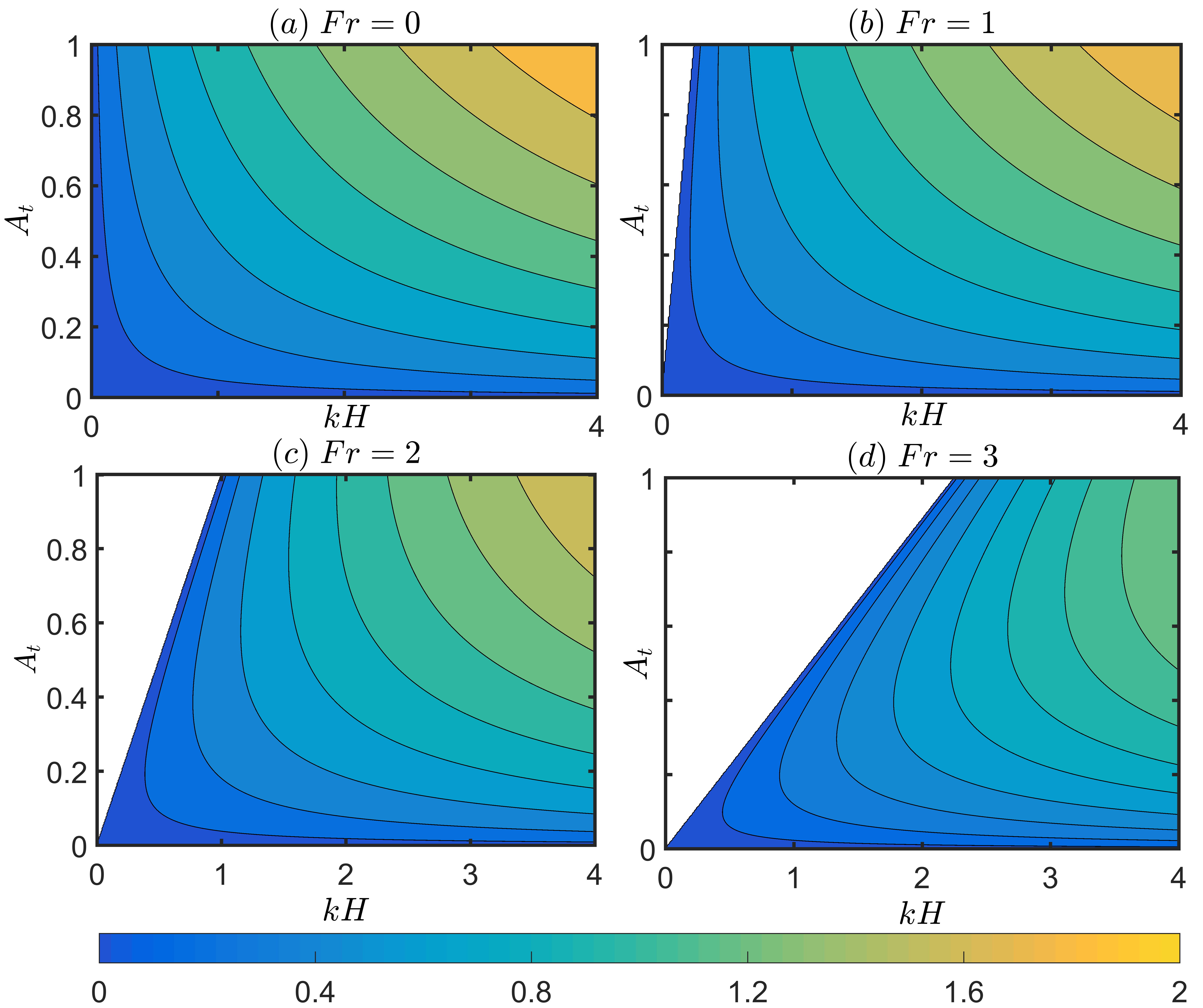}
  \caption{Stability diagram showing non-dimensional growth rate ${\gamma}/{\sqrt{g/H}}$ variation in the $A_t-kH$ plane for (a) $Fr=0$ (b) $Fr=1$ (c) $Fr=2$ (d) $Fr=3$. Here, `H' is an arbitrarily chosen length scale.}
  \label{fig:2}
\end{figure}

Substituting the normal mode perturbations of the form $\exp{\left[\ii (kx-\omega t)\right]}$, where $k$ is the real wavenumber and $\omega$ is the complex frequency, in \eqref{eq:KBC} and \eqref{eq:DBC} and using the procedure  outlined in  ref.\  \cite{drazin2004hydrodynamic}, we obtain the dispersion relation for the system:
\begin{equation}\label{eq:dispersion}
\omega^2-\left(\frac{\rho_1\Omega_1-\rho_2\Omega_2}{\rho_1+\rho_2}\right)\omega+gk\left(\frac{\rho_1-\rho_2}{\rho_1+\rho_2}\right)=0.
\end{equation}
The expression for non-Boussinesq growth rate (i.e. the imaginary part of complex frequency) of RTI in deep water limit is found to be
\begin{align}
\gamma=\sqrt{\left(\frac{\rho_1-\rho_2}{\rho_1+\rho_2}\right)gk-\left(\frac{\rho_1\Omega_1-\rho_2\Omega_2}{2\rho_1+2\rho_2}\right)^2}.
\end{align}
It might be noted here that under the Boussinesq approximation, the effect of density in the (stabilizing) shear term is lost, and we would have simply obtained
\begin{align}
\gamma=\sqrt{\left(\frac{\rho_1-\rho_2}{\rho_1+\rho_2}\right)gk-\left(\frac{\Omega_1-\Omega_2}{2}\right)^2}.
\end{align}
This also means that the Boussinesq approximation will not be able to capture the effect of shear if the shear is uniform i.e. if $\Omega_1=\Omega_2=\Omega$, whereas for non-Boussinesq case and a constant shear we will obtain
\begin{align}
\gamma=\sqrt{A_tgk-\left(\frac{A_t\Omega}{2}\right)^2}.
\label{eq:gamma_main}
\end{align}
It is evident from the above  relation that presence of a uniform shear $\Omega$ always causes a suppression of the growth rate for a single interface system and this effect is prominent at lower wavenumbers. The growth rate plot for RTI in the presence of a uniform shear has been plotted in fig.\ \ref{fig:2}. The value of shear (characterized by Froude number, $Fr=\Omega/\sqrt{g/H})$ has been varied and it can be seen that increasing the shear increases the stable region in the $A_t-kH$ plane. In the absence of any length scale, there is no physical significance of `$H$' and it can be chosen arbitrarily. Our result can be contrasted with that of   ref.\  \cite{zhang2005effect}, in which the expression for RTI growth rate  in presence of velocity discontinuity (and absence of magnetic field) is given by
\begin{align}
\gamma=\sqrt{A_tgk+k^2(1-A_t^2)(\Delta U)^2},
\label{Zhang_eq}
\end{align}
where $\Delta U$ the velocity jump across the interface. It can be seen from \eqref{Zhang_eq} that increasing shear $\Delta U$ increases the instability but from our results, it is evident that increasing shear $\Omega$ actually stabilizes the flow. The contradiction here is due to treating shear as a gradient of velocity  $(d\bar{u}/dz)$ in our case and a discrete velocity jump $\Delta U$ in  ref.\  \cite{zhang2005effect}. The role of a velocity jump (or destabilizing role of shear) will be discussed in the later half of this letter but for now, we have theoretically established that presence of a uniform shear will stabilize RTI. Therefore, the system no longer remains unstable for all values of $k$ and we obtain a cut-off wavenumber below which the system remains stable. The instability condition thus becomes $k>A_t \Omega^2/(4g)$.

Another intriguing and non-intuitive result which comes up exclusively when not making the Boussinesq approximation is the variation of growth rate with the Atwood number, $A_t$. We find  that the variation of growth rate with $A_t$ no longer remains monotonic owing to the presence of $A_t$ in the shear term  as well. Thus the instability, which is driven by the presence of $A_t$ in the `gravity term', is also suppressed by the $A_t$ in the `shear term'.
For a given `$k$' and $\Omega$, the maximum growth rate is obtained at $A_t=2gk/\Omega^2$, and further increasing $A_t$ decreases the growth rate. Although a similar result was obtained in  ref.\  \cite{zhang2005effect}, it was argued that the non-monotonic variation of Atwood number and growth rate was due to the instability being governed by shear at higher Froude numbers. However, we see here that the shear doesn't actually destabilize the flow but rather the stabilizing effect of shear increases with an increase in $A_t$, causing the RTI to stabilize.
It might also be noted from \eqref{eq:dispersion} that for $\rho_1>\rho_2$, the product of roots of the given quadratic equation is positive;  hence if both roots of \eqref{eq:dispersion} are real, then they must be of the same sign. Physically this means that for sufficiently lower wavenumbers for which the system is stable, we obtain two stable waves propagating in the same direction. For a positive $\Omega$, both of the waves (at lower wavenumbers) will be positively propagating. This result will be used later in this letter.

The uniform shear velocity profile (velocity having constant slope) considered here may be too simplistic for many realistic systems, where the characteristics of shear is usually captured by a `shear layer'. The latter will be focused in the subsequent paragraphs. Nevertheless, velocity profile of constant slope is encountered in rotating inviscid fluids  \citep{tao2013nonlinear}. The uniform shear set-up mathematically resembles that of   ref.\  \cite{tao2013nonlinear} in many aspects, where the authors study the effect of rotation on RTI, the axis of rotation being normal to the acceleration of the interface. The dynamic boundary condition in our case, i.e.\ \eqref{eq:DBC}, is similar to that obtained by   ref.\  \cite{tao2013nonlinear} (see their Eq.\, (4)), where `$\Omega\psi$' like term appears due to Coriolis effects. Therefore, the  observation  in  ref.\  \cite{tao2013nonlinear} that Coriolis force diminishes RTI growth rate is in line with ours - uniform shear inhibits RTI. 



\begin{figure}
\centering\includegraphics[width=80mm]{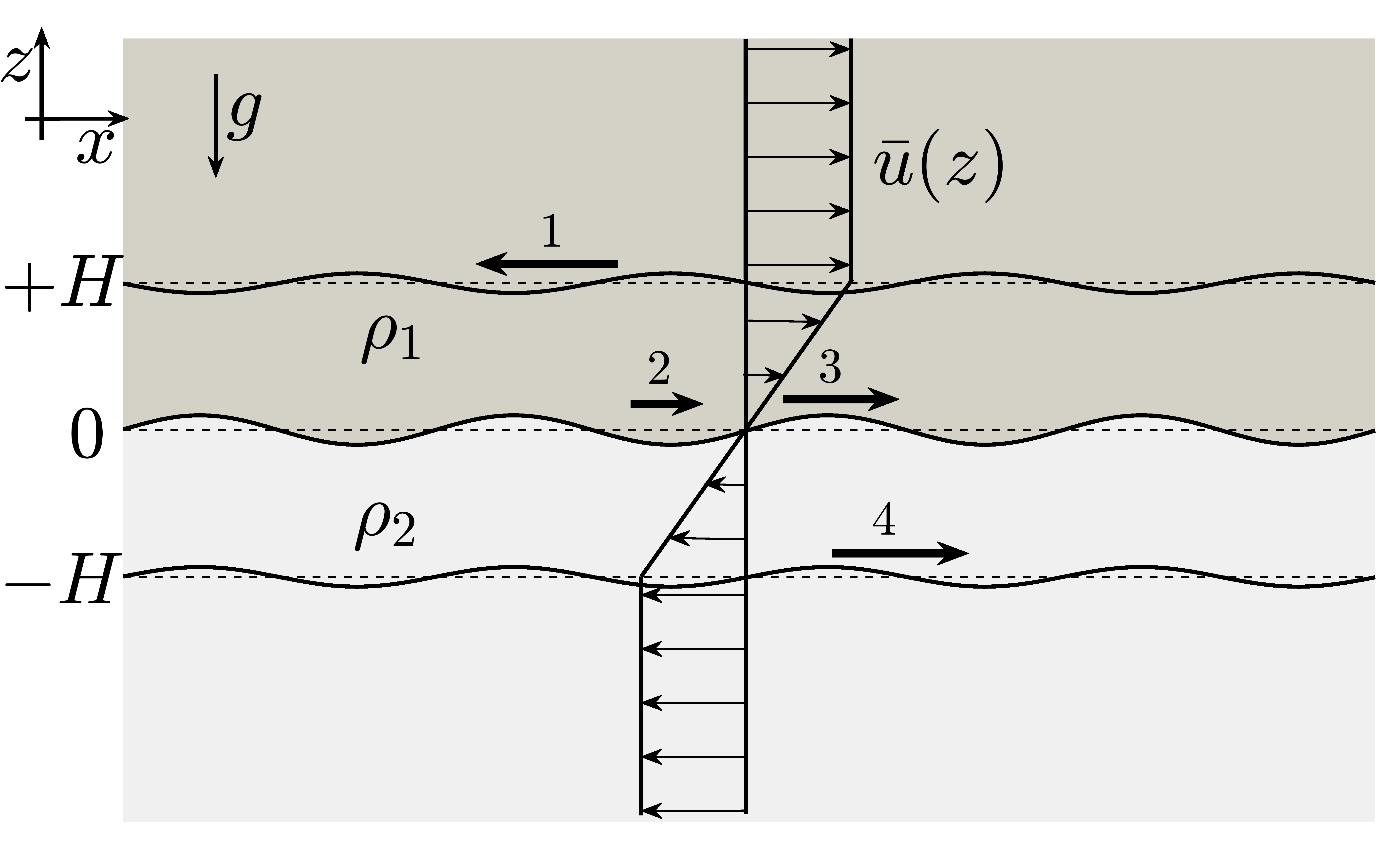}
  \caption{Schematic of  adversely stratified density interface embedded in a shear layer}
  \label{fig:3}
\end{figure}

As mentioned previously, the more realistic setup would be an unstable density interface sandwiched inside a `shear layer'\footnote{Even when shear is not imposed on an unstable density interface, RTI generates a shear layer during the nonlinear stages.}, see fig.\ \ref{fig:3}.
This set-up resembles the Holmboe instability set-up \cite{baines1994mechanism}, except that now the density interface has an unstable stratification. The system is described as follows:
\begin{equation}
\bar{\rho}(z) = \left\{
        \begin{array}{cc}
        \rho_1 & \quad 0 < z \\ 
        \rho_2 & \quad  z < 0
        \end{array}
    \right.
   \quad \Omega(z) = \left\{
        \begin{array}{cc}
        0& \quad H < z \\ 
        \Omega & \quad  -H < z<H \\ 
        0 & \quad z<-H.
        \end{array}
    \right.
\end{equation}

This system is inherently different from the previous one. This is because here multiple interfaces are present; each interface can support one or more waves (marked by `$1$' to `$4$' in fig.\ \ref{fig:3}),  which implies that waves present at different interfaces can interact among themselves and lead to instability \citep{guha2014wave}. While waves `$1$' and `$4$' are vorticity waves (Rossby edge waves if the reference frame is rotating) and their physics is well known \cite{guha2014wave}, waves `$2$' (the slower wave) and `$3$' (the faster wave) are virtually unknown. Existence of a wave very similar to waves `$2$' and `$3$'  has been recently proposed by us \citep{guha2017waves}, where it has been referred to as the ``shear-density wave''. This ``shear-density wave'' appears when a uniform shear is imposed over a density interface (in the absence of gravity). In this case, these two waves are shear-density waves modified by gravity. We note here that the  shear-density waves modified by gravity were also obtained in the set-up with uniform shear. These were the two stably propagating waves at the density interface, which were obtained for lower wavenumbers. These waves should not be mistaken as interfacial gravity waves (which would have propagated in opposite directions) since stratification being adverse, gravity cannot provide the restoring force. 

\begin{figure}
\includegraphics[width=85mm]{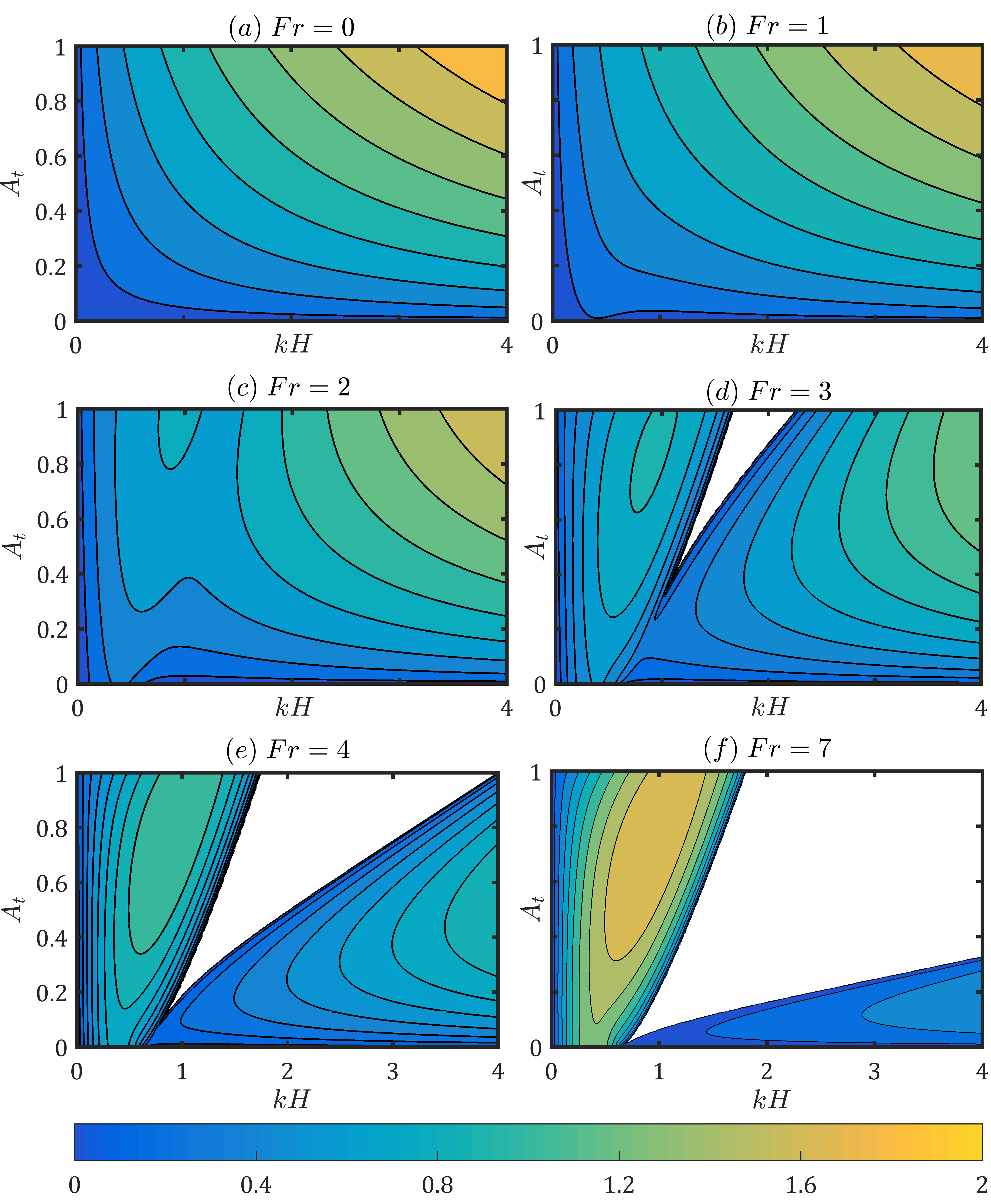}
\caption{Stability diagram for RTI embedded in a shear layer. Two distinct instability regions are observed for high $Fr$, the right one being RTI while the left one is a new kind of shear instability, see text.}
\label{fig:4}
\end{figure}

The non-dimensional growth rate $(\gamma/\sqrt{g/H})$ has been plotted w.r.t. $kH$ in fig.\ \ref{fig:4} for different Froude numbers, $Fr$. Two distinct instability branches are observed for higher Froude numbers (see fig.\ \ref{fig:4}): (i) RTI at $z=0$ due to the adverse buoyancy, and (ii) a new kind of shear instability arising from the interaction between waves `$1$' and`$3$' (and not wave `$2$', which we have confirmed from the dispersion diagram). These two waves form a counter-propagating configuration  \cite{guha2014wave,guha2017waves},  which is necessary for them to lock in phase and grow exponentially. 

The existence of the above-mentioned shear instability has not been reported before in the literature. In fact, the only shear instability arising from the configuration in fig.\ \ref{fig:3} was thought to be Kelvin--Helmholtz instability  \citep{ye2011competitions}(KHI, or more appropriately, instability of Rayleigh's velocity profile). This would have been the case had the interaction been between wave `$1$' and wave `$4$'. However, it is known \citep{ye2011competitions} that KHI is stabilized if $A_t$ is increased, which is clearly not the case here. As can be seen from  fig.\ \ref{fig:4}(e),  $A_t$ stabilizes the shear stability only at very low wavenumbers (which is attributed to KHI).  Even for slightly higher wavenumbers, increasing $A_t$ destabilizes the shear instability even further. 

\begin{figure}[t]
\centering\includegraphics[width=85mm]{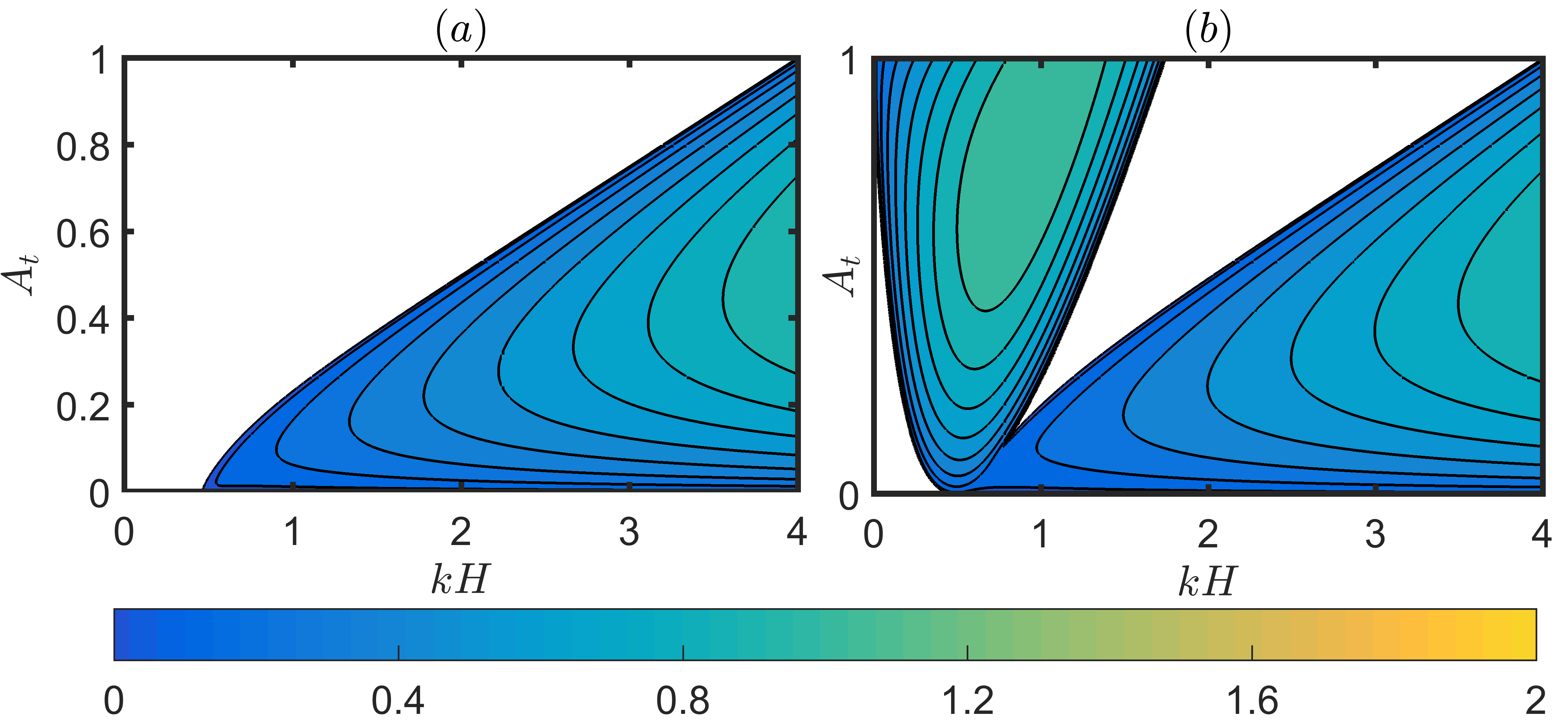}
  \caption{Stability diagram for $Fr=4$, (a) in the absence of wave `1' (b) in the absence of wave `4'.}
  \label{fig:5}
\end{figure}

To conclusively show that KHI does not play a significant role in the set-up under consideration, it is important to remove one of the vorticity waves at a time and see the consequence. First, we remove wave `$1$' by using the following density and shear profiles:     
\begin{equation}
\bar{\rho}(z) = \left\{
        \begin{array}{cc}
        \rho_1 & \quad 0 < z \\ 
        \rho_2 & \quad  z < 0
        \end{array}
    \right.
   \qquad \Omega(z) = \left\{
        \begin{array}{cc}
        \Omega & \quad H < z \\ 
        \Omega & \quad  -H < z<H \\ 
        0 & \quad z<-H
        \end{array}
    \right.
\end{equation}
Since jump in base shear is eliminated at $z=H$, vorticity wave `$1$' disappears. From fig.\ \ref{fig:5}(a) we see that removing wave `$1$' completely removes the region of shear instability. This shows that wave `$1$'  plays a major role in the instability mechanism. Next, we eliminate wave `$4$' (wave `$1$' being present) by eliminating shear jump at $z=-H$, i.e.\ using the following profile:
\begin{equation}
\rho(z) = \left\{
        \begin{array}{cc}
        \rho_1 & \quad 0 < z \\ 
        \rho_2 & \quad  z < 0
        \end{array}
    \right.
   \qquad \Omega(z) = \left\{
        \begin{array}{cc}
        0 & \quad H < z \\ 
        \Omega & \quad  -H < z<H \\ 
        \Omega & \quad z<-H.
        \end{array}
    \right.
\end{equation}
The corresponding growth rate plot in fig.\ \ref{fig:5}(b) shows that very little difference exists with that of fig.\ \ref{fig:4}(e), where wave `$4$' was present. The only small difference appearing at smaller values of $kH$  and $A_t$ is attributed to the presence of KHI in fig.\ \ref{fig:4}(e), which results from the interaction between waves `$1$' and `$4$'.

In summary, presence of simple uniform shear suppresses RTI and this effect increases with increasing shear, characterized here by $Fr$. In presence of shear, the instability does not vary monotonically with stratification (characterized by $A_t$), and is suppressed at higher values of $A_t$. This apparently  counter-intuitive result, mostly prominent at lower wavenumbers, is due to the presence of $A_t$ in the shear term; see (\ref{eq:gamma_main}). Waves with wavenumbers satisfying $k<A_t \Omega^2/(4g)$ become stable.  
Further, a more practically relevant set-up, where an unstable density interface is embedded in a piecewise linear shear layer, is considered. The growth rate contours for the RTI region remains almost the same as that in the case of a uniform shear. Although shear layer configuration stabilizes  RTI similar to the uniform shear case, additional unstable shear instability modes are introduced in the low wavenumber regime.   We show that this shear instability is entirely new and has not been reported in the literature. 
KHI, which also develops from the shear layer, and previously thought to be of key importance, plays a minor role. This implies that the billows observed in the nonlinear stages of RTI, which are traditionally attributed to secondary KHI  \cite{he1999lattice}, may indeed be the nonlinear manifestation of this new  shear instability.

 \bibliography{paper1}
\end{document}